\documentclass{ws-ijmpe}
\usepackage[super,compress]{cite}

\begin{document}

\markboth{M. Thoennessen}{2018 Update of the Discoveries of Isotopes}

\catchline{}{}{}{}{}

\title{2018 update of the discoveries of nuclides}

\author{\footnotesize M. Thoennessen}

\address{American Physical Society \\
Ridge, NY 11961, USA\\  \vspace*{0.2cm}
National Superconducting Cyclotron Laboratory and \\
Department of Physics \& Astronomy \\
Michigan State University\\
East Lansing, Michigan 48824, USA\\
thoennessen@nscl.msu.edu}

\maketitle

\begin{history}
\received{Day Month Year}
\revised{Day Month Year}
\end{history}

\begin{abstract}
The 2018 update of the discovery of nuclide project is presented. 50 new nuclides were observed for the first time in 2018. A large number of isotopes is still only published in conference proceedings or internal reports. 
\end{abstract}


\ccode{PACS numbers: 21.10.-k, 29.87.+g}


\section{Introduction}

This is the sixth update of the isotope discovery project which was originally published in a series of papers in Atomic Data and Nuclear Data Tables from 2009 through 2013 (see for example the first \cite{2009Gin01} and last \cite{2013Fry01} papers). Two summary papers were published in 2012 and 2013 in Nuclear Physics News \cite{2012Tho03} and Reports on Progress in Physics \cite{2013Tho02}, respectively, followed by annual updates in 2014 \cite{2014Tho01}, 2015 \cite{2015Tho01}, 2016 \cite{2016Tho02}, 2017 \cite{2017Tho01}, and 2018 \cite{2018Tho01}. In  2016 a description of the discoveries from a historical perspective was published in the book ``The Discovery of Isotopes -- A complete Compilation''  \cite{2016Tho01}.

\section{New discoveries in 2018}
\label{New2018}

In 2018, the discoveries of 50 new nuclides were reported in refereed journals. Most of the new isotopes were very neutron-rich nuclides (43) which were all discovered at RIKEN in Japan. In addition, five nuclides along the proton-dripline and two transuranium nuclides were newly observed. Five of the nuclides are beyond the proton (three) and neutron (two) dripline. Table \ref{2018Isotopes} lists details of the discoveries including the production methods. 

\begin{table}[pt]
\tbl{New nuclides reported in 2018. The nuclides are listed with the first author, submission date, and reference of the publication, the laboratory where the experiment was performed, and the production method (PF = projectile fragmentation, FE = fusion evaporation, SB = secondary beams). \label{2018Isotopes}}
{\begin{tabular}{@{}llrclc@{}} \toprule 
Nuclide(s) & First Author & Subm. Date & Ref. & Laboratory & Type \\ \colrule
$^{161}$Pr , $^{163}$Nd, $^{164}$Pm, $^{165}$Pm, & N. Fukuda & 2/20/2017  &\refcite{2018Fuk01}& RIKEN & PF \\
$^{167}$Sm, $^{169}$Eu, $^{171}$Gd, $^{173}$Tb, & & & & & \\
$^{174}$Tb, $^{175}$Dy, $^{176}$Dy, $^{177}$Ho, & & & & & \\
$^{178}$Ho, $^{179}$Er, $^{180}$Er & & & & & \\
$^{104}$Rb, $^{113}$Zr, $^{116}$Nb, $^{119}$Mo, & Y. Shimizu  & 3/8/2017 &\refcite{2018Shi01}& RIKEN & PF \\
$^{122}$Tc, $^{125}$Ru, $^{128}$Rh, $^{130}$Pd, & & & & & \\
$^{131}$Pd, $^{140}$Sn, $^{142}$Sb, $^{145}$Te, & & & & & \\
$^{146}$I, $^{147}$I, $^{149}$Xe, $^{150}$Xe, $^{157}$La& & & & & \\
$^{219}$Np  & H.B. Yang & 9/18/2017 & \refcite{2018Yan01}& Lanzhou & FE \\
$^{224}$Np  & T.H. Huang & 1/10/2018 & \refcite{2018Hua01}& Lanzhou & FE \\
$^{47}$P, $^{49}$S, $^{52}$Cl, $^{54}$Ar, $^{57}$K, & O.B. Tarasov & 5/7/2018 & \refcite{2018Tar01}& RIKEN & PF \\
$^{59}$Ca, $^{60}$Ca, $^{62}$Sc, $^{59}$K & & & & & \\
$^{28}$Cl, $^{30}$Cl , $^{29}$Ar & I. Mukha & 6/9/2018 & \refcite{2018Muk01}& GSI & SB \\
$^{104}$Te, $^{108}$Xe & K. Auranen & 7/31/2018 & \refcite{2018Aur01}& Argonne & FE \\
$^{20}$B, $^{21}$B	 & S. Leblond & 9/7/2018 & \refcite{2018Leb01} & RIKEN & SB \\ 
\botrule
\end{tabular}}
\end{table}

Fukuda et al. discovered 15 new nuclides in ``Identification of new neutron-rich isotopes in the rare-earth region produced by 345 MeV/nucleon $^{238}$U''\cite{2018Fuk01}. A 345 MeV/nucleon $^{238}$U  beam from the  RIKEN RIBF accelerator complex bombarded beryllium targets and in-flight fission fragments were separated with the BigRIPS separator. The two-stage isotope separation mode was used and the nuclides were identified using the $\Delta$E-TOF-B$\rho$ method. ``We observed a total of 29 new neutron-rich isotopes: $^{153}$Ba, $^{154,155,156}$La, $^{156,157,158}$Ce, $^{156,157,158,159,160,161}$Pr, $^{162,163}$Nd, $^{164,165}$Pm, $^{166,167}$Sm, $^{169}$Eu, $^{171}$Gd, $^{173,174}$Tb, $^{175,176}$Dy, $^{177,178}$Ho, and $^{179,180}$Er.'' However, Fukuda et al. is only credited with the discovery of the 15 isotopes listed in Table \ref{2018Isotopes} as the others had previously been reported by Wu \cite{2017Wu01}. Unfortunately, Fukuda et al. did neither acknowledge nor reference the earlier paper.

In the paper ``Observation of new neutron-rich isotopes among fission fragments from in-flight fission of 345 MeV/nucleon $^{238}$U: Search for new isotopes conducted concurrently with decay measurement campaigns'' Shimizu et al. reported the discovery of 17 new nuclides\cite{2018Shi01}. The RIKEN RIBF accelerator complex was used to bombard a 2.92-mm-thick beryllium target with a 345 MeV/nucleon $^{238}$U beam. Fission fragments were then identified following the BigRIPS and the ZeroDegree spectrometer. The subsequent decays were recorded withe EURICA setup. ``In total, we have produced and identified 36 new neutron-rich isotopes: $^{104}$Rb, $^{113}$Zr, $^{116}$Nb, $^{118,119}$Mo, $^{121,122}$Tc, $^{125}$Ru, $^{127,128}$Rh, $^{129,130,131}$Pd, $^{132}$Ag, $^{134}$Cd, $^{136,137}$In, $^{139,140}$Sn, $^{141,142}$Sb, $^{144,145}$Te, $^{146,147}$I, $^{149,150}$Xe, $^{149,150,151}$Cs, $^{153,154}$Ba, and $^{154,155,156,157}$La.'' Only the 17 isotopes listed in Table \ref{2018Isotopes} are credited to Shimizu et al. because the others had previously been published by Lorusso et al. ($^{118}$Mo, $^{121}$Tc, $^{127}$Rh, $^{129}$Pd, $^{132}$Ag, $^{134}$Cd, $^{136,137}$In, $^{139}$Sn, $^{141}$Sb, and $^{144}$Te)\cite{2015Lor01} and Wu et al. ($^{149,150,151}$Cs, $^{153,154}$Ba, and $^{154,155,156}$La)\cite{2017Wu01}. While Shimizu et al. mentioned the Lorusso et al. paper: ``Note that $\beta$-decay half-life measurements have already been reported for some of the new isotopes observed with relatively high production yields.'' they did neither acknowledge nor reference Wu et al.

The first identification of $^{219}$Np was reported by Yang et al. in the paper ``Alpha decay properties of the semi-magic nucleus $^{219}$Np'' \cite{2018Yan01}. A 191.5 MeV $^{36}$Ar was accelerated with the Sector Focusing Cyclotron of the heavy ion research facility in Lanzhou (HIRFL), China. Residues from the fusion-evaporation reaction $^{187}$Re($^{36}$Ar,4n) were identified with the gas-filled recoil separator SHANS. ``According to the observed $\alpha$-decay chain, an energy of E$_\alpha$ = 9039(40) keV and a half-life of T$_{1/2}$ = 0.15$^{+0.72}_{-0.07}$ ms were determined for $^{219}$Np.'' Previously, Devaraja et al. had reported $^{219}$Np in the decay chain starting at $^{223}$Am, however, neither the $\alpha$-decay energy nor the lifetime were measured\cite{2015Dev01}.

Huang et al. reported the discovery of $^{224}$Np in the paper ``Identification of the new isotope $^{224}$Np'' \cite{2018Hua01}. The Sector-Focusing Cyclotron (SFC) of the Heavy Ion Research Facility in Lanzhou (HIRFL) accelerated a $^{40}$Ar beam to 188 MeV which then bombarded a 460 $\mu$g/cm$^2$ thick $^{187}$Re target. The gas-filled recoil separator, Spectrometer for Heavy Atoms and Nuclear Structure (SHANS) selected $^{224}$Np residues from the fusion-evaporation reaction $^{187}$Re($^{40}$Ar,3n). The isotopes were implanted into a double-sided silicon strip detector which also recorded subsequent $\alpha$-decays. ``Two $\alpha$-decay branches of $^{224}$Np were identified through spatial and temporal correlation measurements, populating two low-lying isomeric states in $^{220}$Pa. Their energies and half-life were determined to be E$_{\alpha 1}$ = 8868(62) keV, E$_{\alpha_2}$ = 9137(20) keV, and T$_{1/2}$ = 38$^{+26}_{-11}$ $\mu$s respectively.''

In the paper ``Discovery of $^{60}$Ca and Implications For the Stability of $^{70}$Ca'', Tarasov et al. described the first observation of nine isotopes \cite{2018Tar01}.  A 345 MeV/u $^{70}$Zn beam from the RIKEN radioactive ion-beam factory (RIBF) accelerator complex irradiated $^9$Be targets. Projectile fragmentation products of interest were separated with the BigRIPS separator and identified event-by-event by the PID(Z,A,q) method. ``The observed fragments include eight new isotopes that are the most neutron-rich nuclides of the elements from phosphorus to scandium, $^{47}$P(12), $^{49}$S(5), $^{52}$Cl(2), $^{54}$Ar(13), $^{57}$K(8), $^{59}$Ca(9), $^{60}$Ca(2), $^{62}$Sc(2) (the number of detected events is given in brackets). One event consistent with $^{59}$K was observed as well.''

The first observation of $^{28}$Cl, $^{30}$Cl and $^{29}$Ar was reported by Mukha et al. in the paper ``Deep excursion beyond the proton dripline. I. Argon and chlorine isotope chains'' \cite{2018Muk01}. A 885 A$\cdot$MeV $^{36}$Ar beam from the SIS-FRS facility at GSI was used to produce a 620 MeV A$\cdot$MeV $^{31}$Ar beam in the first half of the FRS set in a separator-spectrometer mode. Unbound reaction products were generated at the $^9$Be secondary target located at the FRS middle focal plane. The projectile-like fragments were analyzed by the second half of the FRS and the decay particles were measured by a double-sided silicon microstrip detector array. $^{29}$Ar was populated by two-neutron knockout reactions, while $^{28}$Cl and $^{30}$Cl were populated either directly in the fragmentation of $^{31}$Ar or by proton emission from the corresponding $^{31,29}$Ar isotopes. ``The ground states of the previously unknown isotopes $^{30}$Cl and $^{28}$Cl have been observed for the first time, providing the 1p-separation energies S$_p$ of $-$0.48(2) and $-$1.60(8), MeV, respectively... The first-time observed state in $^{29}$Ar with S$_{2p}$ = $-$5.50(18) MeV can be identified as either a ground state or an excited state according to different
systematics.''

Auranen et al. described the first identification of $^{104}$Te and $^{108}$Xe in ``Superallowed $\alpha$ Decay to Doubly Magic $^{100}$Sn'' \cite{2018Aur01}. The ATLAS facility of Argonne National Laboratory was used to accelerate $^{58}$Ni to 245 MeV and bombard 450 $\mu$g/cm$^2$ $^{54}$Fe targets. Evaporation residues from the reaction 
$^{54}$Fe($^{58}$Ni,4n)$^{108}$Xe were separated with the Fragment Mass Analyzer (FMA) and implanted in a double-sided silicon strip detector (DSSD). Subsequent $\alpha$ decays were measured in the DSSD and a box of eight single-sided strip detectors. ``The $\alpha$-particle energy reconstruction resulted in values of E$_\alpha$($^{108}$Xe) = 4.4(2) MeV and E$_\alpha$ =$^{104}$Te = 4.9(2) MeV.'' The observation of $^{104}$Te had previously been reported by Celikovic et al. in an unpublished Ph.D. Thesis in 2013 \cite{2013Cel01}.

The discovery of $^{20}$B and $^{21}$B was reported in the paper ``First Observation of $^{20}$B and $^{21}$B'' by Leblond et al. \cite{2018Leb01}. A 345 MeV/nucleon $^{48}$Ca beam from the Radioactive Isotope Beam Factory (RIBF) of the RIKEN Nishina Center irradiated a 20 mm thick beryllium target to produce secondary beams of $^{22}$N and $^{22}$C which were separated with the BigRIPS fragment separator. These isotopes impinged on a secondary 1.8 g/cm$^2$ carbon target located at the SAMURAI spectrometer. Unbound resonant states were then reconstructed from fragments detected behind the SAMURAI superconducting dipole magnet and neutrons measured in the NEBULA neutron array. ``Two-proton removal from $^{22}$N populated a prominent resonancelike structure in $^{20}$B at around 2.5 MeV above the one-neutron decay threshold, which is interpreted as arising from the closely spaced 1$^-$,2$^-$ ground-state doublet predicted by the shell model. In the case of proton removal from $^{22}$C, the $^{19}$B plus one- and two-neutron channels were consistent with the population of a resonance in $^{21}$B 2.47 $\pm$ 0.19 MeV above the two-neutron decay threshold, which is found to exhibit direct two-neutron decay.''

\section{Status at the end of 2018}

The 50 new discoveries in 2018 increased the total number of observed isotopes to 3302. It corresponds to the most isotopes discovered in a year since 2012 when 68 new isotopes were reported for the first time. These discoveries were reported by 916 different first authors in 1551 papers and a total of 3762 different co-authors. This year there were no changes to any of the assignments from previous years. 

In the ranking of countries with the most discoveries, Japan passed France to enter the top five. In the list of co-authors, T. Kubo (RIKEN/MSU) jumped from 6$^{th}$ to 4$^{th}$ place with 213 isotopes discovered, while O. Tarasov (MSU) moved from 16$^{th}$ to 11$^{th}$ in the list of first-author publications with 33 isotopes discovered. Also, there are now more new isotopes reported in Physical Review Letters (268) than in Nature (261) ranking 5$^{th}$ and 6$^{th}$ in the journal category, respectively. Further statistics can be found on the discovery project website \cite{2011Tho03}.

\begin{figure}[pt] 
\centerline{\psfig{file=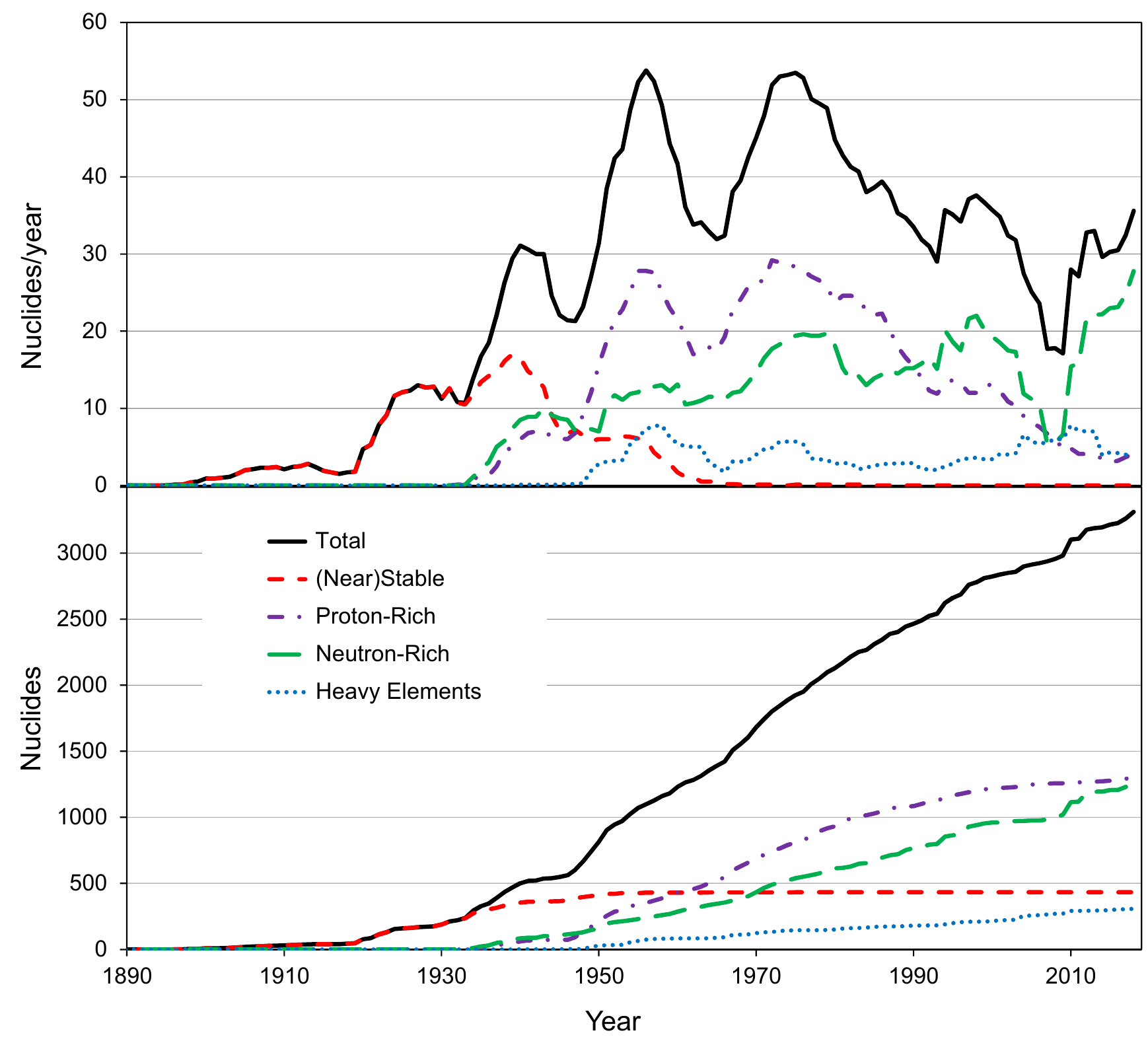,width=12.4cm}}
\caption{Discovery of nuclides as a function of year. The top figure shows the 10-year running average of the number of nuclides discovered per year while the bottom figure shows the cumulative number.  The total number of nuclides shown by the black, solid lines are plotted separately for near-stable (red, short-dashed lines), neutron-deficient (purple, dot-dashed lines), neutron-rich (green, long-dashed lines) and transuranium (blue, dotted lines) nuclides. This figure was originally published in Ref. 5 and updated to include the most recent data. \label{f:timeline} }
\vspace*{-0.1cm}
\end{figure}

Figure \ref{f:timeline} shows the current status of the evolution of the nuclide discoveries for four broad areas of the nuclear chart, (near)stable, proton-rich, neutron-rich, and the region of the heavy elements. The figure was adapted  from the 2014 review\cite{2014Tho01} and was extended to include all isotopes discovered until the end of 2018. The top part of the figure shows the ten-year average of the number of nuclides discovered per year while the bottom panel shows the integral number of nuclides discovered. 

The current overall ten-year average rate is the highest since almost twenty years (2000) with 35.7 nuclide discoveries per year. The ten-year average for neutron-rich isotopes keeps increasing reaching another all-time high of 27.8. However, overall there are still more known proton-rich (1298) than neutron-rich nuclides (1274), although the difference is now only twenty-four isotopes. If the observation of new isotopes at RIKEN tentatively reported in conference proceedings (See next section) are published (more than 30) the number of known neutron-rich isotopes will be larger than the number of known proton-rich nuclides for the first time. 

Also, the ten-year average for transuranium isotopes dropped to its lowest value (3.7) since 2000 (3.4) falling below the rate for proton-rich nuclides (4.1) for the first time since 2008. Since 2010, when the second largest number of new transuranium isotopes were reported (21), less than five isotopes were discovered per year.

\section{Discoveries not yet published in refereed journals}

Table \ref{reports} lists the isotopes which so far still have only been presented in conference proceedings or internal reports. Eleven isotopes that were listed in last year's corresponding table have been added to the refereed literature this year including $^{20}$B published in the last week of the year by Leblond et al.\cite{2018Leb01} and the nine neutron-rich nuclides in the calcium region discovered by Tarasov et al. at RIKEN \cite{2018Tar01}. $^{104}$Te was listed as reported in the 2013 Ph.D. Thesis by Celikovic\cite{2013Cel01} but was discovered this year by Auranen et al. \cite{2018Aur01}. It should be mentioned that 15 isotopes reported by Fukuda et al.\cite{2018Fuk01} and 17 isotopes reported by Shimizu et al. \cite{2018Shi01} at the beginning of the year had not been included in the list of ``not yet pubished'' isotopes.

The table contains five new isotopes: $^{24}$N and $^{25}$N were the subject of the Ph.D. thesis of Deshayes \cite{2018Des01}, $^{11}$O was posted on the arXives by Webb et al. \cite{2018Web01}, and $^{79}$Co and $^{84}$Cu were reported in the most recent RIKEN Annual Report \cite{2018Shi02}.

An additional three isotopes were presented at conferences and mentioned in the corresponding abstracts. Kondo \cite{2018Kon01}  reported the latest results from the SAMURAI21 collaboration for $^{27}$O and $^{28}$O at DREB2018, the 10$^{th}$ International Conference on Direct Reactions with Exotic Beams, and Shimada \cite{2018Shi03} presented the first spectroscopy of the unbound neutron-rich nucleus $^{30}$F at the 5$^{th}$ Joint Meeting of the APS Division of Nuclear Physics and the Physical Society of Japan. It should be mentioned that $^{24}$N and $^{25}$N mentioned above and listed in the table were also presented at DREB2018 by Orr \cite{2018Orr01}.

\begin{table}[t]
\tbl{Nuclides only reported in proceedings or internal reports until the end of 2018. The nuclide, first author, reference and year of proceeding or report are listed. \label{reports}}
{\begin{tabular}{@{}llcc@{}} \toprule
\parbox[t]{6.8cm}{\raggedright Nuclide(s) } & \parbox[t]{2.3cm}{\raggedright First Author} & Ref. & Year \\ \colrule
$^{21}$C		&	 S. Leblond 	&	\refcite{2015Leb01},\refcite{2015Leb02}	&	2015	 \\ 
			&	 N. A. Orr 	&	\refcite{2016Orr01}	&	2016	 \\ 
$^{24}$N, $^{25}$N		&	 Q. Deshayes 	&	\refcite{2018Des01}	&	2018	 \\ 
$^{11}$O		&	 T. B. Webb 	&	\refcite{2018Web01} &	2018	 \\ 
$^{39}$Na	& D. S. Ahn 	&	\refcite{2016Ahn01},\refcite{2018Ahn01}	&	2016/18	 \\ 
			& O. B. Tarasov 	&	\refcite{2017Tar01}	&	2017	 \\ 
$^{79}$Co, $^{84}$Cu & Y. Shimizu & \refcite{2018Shi02} & 2018 \\
$^{86}$Zn$^a$, $^{88}$Ga, $^{89}$Ga, $^{91}$Ge, $^{93}$As$^a$, $^{94}$As, $^{96}$Se, $^{97}$Se	&	 Y. Shimizu 	&	\refcite{2015Shi01}	&	2015	 \\ 
$^{99}$Br, $^{100}$Br & & & \\
$^{98}$Sn	&	  I. Celikovic 	&	\refcite{2013Cel01}	&	2013	 \\ 
$^{155}$Ba, $^{159}$Ce, $^{164}$Nd, $^{166}$Pm, $^{168}$Sm,  	&	 N. Fukuda 	&	\refcite{2015Fuk01}	&	2015	 \\
$^{170}$Eu, $^{172}$Gd, $^{173}$Gd, $^{175}$Tb, $^{177}$Dy, $^{178}$Ho, $^{179}$Ho,& & & \\
 $^{180}$Er, $^{181}$Er, $^{182}$Tm, $^{183}$Tm & & & \\
$^{126}$Nd, $^{136}$Gd, $^{138}$Tb, $^{143}$Ho$^b$, $^{150}$Yb, $^{153}$Hf	&	 G. A. Souliotis 	&	\refcite{2000Sou01}	&	2000	 \\
	$^{143}$Er, $^{144}$Tm	&	 R. Grzywacz 	&	\refcite{2005Grz01}	&	2005	 \\
	& K. Rykaczewski & \refcite{2005Ryk01} & 2005 \\
	& C. R. Bingham & \refcite{2005Bin01} & 2005 \\
$^{230}$At, $^{232}$Rn	&	 J. Benlliure 	&	\refcite{2010Ben02}	&	2010	 \\
					&	  	&	\refcite{2015Ben01}	&	2015	 \\
$^{235}$Cm	&	 J. Khuyagbaatar 	&	\refcite{2007Khu01}	&	2007	 \\
$^{252}$Bk, $^{253}$Bk	&	 S. A. Kreek 	&	\refcite{1992Kre01}	&	1992	 \\
$^{262}$No 	&	 R. W. Lougheed 	&	\refcite{1988Lou01},\refcite{1989Lou01}	&	1988/89	 \\
	&	 E. K. Hulet 	&	\refcite{1989Hul01}	&	1989	 \\
$^{261}$Lr, $^{262}$Lr	&	 R. W. Lougheed 	&	\refcite{1987Lou01}	&	1987	 \\
	&	 E. K. Hulet 	&	\refcite{1989Hul01}	&	1989	 \\
	&	 R. A. Henderson 	&	\refcite{1991Hen01}	&	1991	 \\
$^{255}$Db	&	 G. N. Flerov	&	\refcite{1976Fle01}	&	1976	 \\
		& A.-P. Lepp\"anen	& \refcite{2005Lep01} & 2005 \\
$^{280}$Ds	&	 K. Morita	&	\refcite{2015Mor01}	&	2014	 \\
\botrule
\vspace*{-0.2cm} & & & \\
$^a$ also published in ref. \refcite{2018Shi02} \\
$^b$ also published in ref. \refcite{2003Sew02} \\
\end{tabular}}
\end{table}

\section{Summary}

Considering that Table \ref{reports} does not differ significantly from last year's table, it is not likely that the recent positve trend in new isotope discoveries will continue in 2019. The medium mass proton-rich nuclides as well as the transuranium nuclides listed in the table were reported eight or more years ago, so that their discoveries probably have to wait for new experiments. The one exception is $^{280}$Ds which was reported in an annual report in 2015 by Morita et al.\cite{2014Mor01}, however, the discovery could not be unambiguously identified in the subsequent refereed publication \cite{2017Kaj01}. Thus, the discovery of $^{280}$Ds also has to wait for a follow-up experiment.

Unless at least some of the 33 neutron-rich nuclides reported from RIKEN will be published this year, it is expected that only a few isotopes will be discovered in 2019. Other than these isotopes there is only one possible candidate listed in the table: $^{11}$O, posted on the arXiv and submitted for publication in December 2018\cite{2018Web01}.

If indeed only a few new isotopes are reported during the present year, another update of the discoveries of nuclides at the end of 2019 will not be warranted. The most current status will continue to be posted on the discovery project website \cite{2011Tho03}. Input, feedback, and comments from researchers are encouraged to ensure that the compilation is always complete and up-to-date.


\bibliographystyle{ws-ijmpe}
\bibliography{isotope-references-etal}

\end{document}